\documentstyle[12pt,epsf]{article}

\renewcommand{\theequation}{\thesection.\arabic{equation}}

\begin{document}
\begin{titlepage}
  \begin{flushright}
    UT-801 \\[-1mm]
    KUNS-1479 \\[-1mm]
    HE(TH)97/19 \\[-1mm]
    hep-ph/9712530
  \end{flushright}

  \begin{center}
    \vspace*{1cm}
    {\Large\bf Quark and Lepton Mass Matrix\\[3mm] in an
    Asymptotically Non-Free Theory}
    \vspace{1.5cm}

    {\large
      Masako {\sc Bando}\footnote{E-mail address: {\tt
          bando@aichi-u.ac.jp}},
      Joe {\sc Sato}\footnote{E-mail address: {\tt
          joe@hep-th.phys.s.u-tokyo.ac.jp}} and
      Koichi {\sc Yoshioka}\footnote{E-mail address: {\tt
          yoshioka@gauge.scphys.kyoto-u.ac.jp}}}
    \vspace{7mm}

    $~^*${\it Aichi University, Aichi 470-02, Japan} \\[1mm]
    $~^{\dagger}${\it Department of Physics, University
      of Tokyo, Tokyo 113, Japan} \\[1mm]
    $~^{\ddagger}${\it Department of Physics, Kyoto
      University, Kyoto 606-01, Japan}
    \vspace{1.5cm}

    \begin{abstract}
      We analyze fermion mass-matrix structure in an asymptotically
      non-free model with $4 + \bar 1$ generations. The texture at the 
      GUT scale is uniquely determined by supposing that the masses of
      heavy up-type quarks (charm as well as top) are realized as
      their infrared fixed point values. By assuming $SO(10)$ GUT-like
      relations for Yukawa couplings in this model, this texture can
      explain all fermion masses and quark mixing with only one small
      parameter, which is almost equal to the Cabibbo angle.
    \end{abstract}
  \end{center}
\end{titlepage}

\newpage

\section{Introduction}
\setcounter{equation}{0}\setcounter{footnote}{0}

Determining the origin of fermion masses and mixing is one of the most 
important problems in constructing matter unification models. In the
framework of the minimal supersymmetric standard model (MSSM), which
is successful in attaining gauge coupling unification \cite{GUT},
there are many works dedicated obtaining fermion mass structure
\cite{mass}. One interesting feature which has been found is that the
MSSM with the $SU(5)$-GUT relation is almost consistent with the
observed bottom-tau mass ratio \cite{b-tau}.

In previous papers \cite{BOST,BSY}, we investigated an extension of
the MSSM with an asymptotically non-free (ANF) property motivated by
the possibility of dynamical gauge bosons. The difference between
dynamical gauge bosons and elementary gauge bosons is the existence of 
the compositeness condition at some scale below which they behave as
if they are asymptotically non-free gauge fields \cite{DGB}. It is
interesting to determine whether or not the present standard gauge
theory with ANF character preserves the above good predictions of the
MSSM\@. As an example we consider the simplest case in which there are
2 extra generations, forming a generation--mirror generation pair (the
4th and anti--4th generations) with $SU(2)_W\times U(1)_Y$ invariant
Dirac masses $M$. There are many works that address the idea of extra
fermions, especially vector-like families, which are motivated by
various physical backgrounds, string models \cite{string}, composite
models \cite{tc}, grand unified models \cite{gum}, etc., and the
possibility of having new families just above the 3-generation
families is interesting. This vector-like matter is compatible with
the constraints determined by the LEP measurements, namely the
so-called Peskin--Takeuchi constraints \cite{P-T}. Following Babu and
Pati in Ref.\ \cite{gum}\@, we call this the extended supersymmetric
standard model (ESSM) hereafter. In the ESSM three gauge couplings are
also unified at almost the same scale as that of the MSSM but with
different unified coupling. An immediate consequence of the ESSM is
that all three gauge couplings of $SU(3)_C, SU(2)_W$ and $ U(1)_Y$,
are ANF: they become larger as they evolve up to coincide at the
unification scale. This model is a typical one in which the
asymptotically non-free gauge couplings remain reasonably small up to
the GUT scale (see Ref.\ \cite{BOST} and references therein) and the
model is worth investigating.

Another interesting feature of the ESSM is the infrared fixed-point
structure of Yukawa couplings due to the ANF gauge couplings. In
Ref.\ \cite{BSY}, we pointed out that the top and bottom quark masses
are reproduced as infrared fixed point (IRFP) predictions, which are
almost insensitive to initial GUT conditions due to the ANF characters
of gauge couplings. Such strong convergence of Yukawa couplings to
their infrared fixed points is a common feature appearing in ANF
theories \cite{FP-ANF}. In such a case, the IRFP structure acts to
translate symmetry structures into quantitative predictions for
low-energy physical parameters, as stressed in Ref.\ \cite{ross}\@. We 
demonstrated how strongly the couplings are focused into their
infrared fixed points in ANF theories as well as the structure of the
renormalization-group flows by comparing the ESSM with the MSSM
\cite{BSY}.

In the above mentioned analyses, we neglected the small Yukawa
couplings of the first and second generations. However, the existence
of Yukawa couplings to the 4th and $\bar 4$th generations may affect
their structure in some way. In this paper we study the fermion masses
of 5 full generations. As is well known, the ordinary fermion masses
show typical hierarchical structures with small parameters $\sim
O(10^{-(1-3)})$, which have to be introduced from the beginning in GUT
textures. In the ESSM case we have one more hierarchical factor; the
ratio of the usual $SU(2)_W\times U(1)_Y$ breaking mass $m$ and the
invariant mass $M$ $(m \ll M)$. By making full use of the infrared
fixed point structure and the hierarchical ratio $m/M$, we shall
determine their textures at the GUT scale.

In section 2 we give a quick summary of the present status of the
fermion masses and mixing and of some features of our previous
analysis in the ESSM\@. In section 3 we analyse the simplest case in
which the invariant mass appears only in the 4th and $\bar 4$th
fermions. It is found that the texture at the GUT scale is uniquely
determined if we impose the condition that the masses of heavy up-type
quarks are realized as their infrared fixed point values, together
with $SO(10)$ GUT-like relations for Yukawa couplings. We will see
that this texture reproduces the fermion masses and quark mixing by
introducing only one small parameter $\epsilon \sim 0.2$, which is
almost equal to the Cabibbo angle. Concluding remarks and comments are
made in section 4. Appendix A is devoted to more complicated analysis
in which more invariant masses $M$ are included or texture is
non-symmetric. We find that none of them can reproduce the present
experimental data of masses and mixing. The renormalization group
equations in the ESSM are given in Appendix B.

\section{Fermion masses and mixing}
\setcounter{equation}{0}\setcounter{footnote}{0}

\subsection{Issues of fermion mass}

The possible sources which determine fermion masses are as follows.
\begin{enumerate}
\item The texture of Yukawa matrix at the GUT scale \cite{texture}: \
  The GUT relations of Yukawa couplings are important to account for
  the hierarchical structure between generations. Although no rational
  basis is known to determine the intergenerational relationship,
  fermion masses seem to exhibit typical hierarchical structures. This
  may indicate the existence of a kind of generation quantum
  numbers. Actually there are many papers to explain hierarchical
  Yukawa couplings by assuming horizontal symmetry \cite{h-sym}, using
  anomalous $U(1)$ \cite{anomalous}, etc., where we can make an active
  use of higher dimensional operators which effectively give very
  small Yukawa couplings. In any case it is important to determine the
  texture of the Yukawa matrix at the GUT scale according to some yet
  unknown rule. We shall determine the texture phenomenologically
  assuming that some unknown mechanism yields hierarchical structure.
\item Running couplings from the GUT scale to weak scale: \
  Once we fix the texture at GUT scale the renormalization group
  equations (RGE) tell us the resultant Yukawa couplings at low-energy
  scale. Since the Yukawa couplings, except for those of third
  generation, are much smaller than the strength of gauge couplings
  (especially the QCD coupling), usually the renormalization effect
  comes mainly from QCD\@. The relative ratio between quark Yukawa
  couplings is not largely changed by RGE\@. The only important factor
  is the ratio of Yukawa couplings of quarks and leptons.
\item Mixing pattern of light Higgses: \
  For the moment we treat the Higgs potential, and Higgs mixing
  (including $\tan \beta$), as free parameters to be determined
  phenomenologically, since we always encounter the well-known
  fine-tuning problem.
\end{enumerate}
For later convenience, we list the masses of the present existing
fermions at $M_Z$ scale \cite{qmass}:
\begin{eqnarray}
  \begin{array}{lll}
    m_u \sim  2.33_{-0.45}^{+0.42} {\rm ~MeV}, &
    m_d \sim  4.69_{-0.66}^{+0.60} {\rm ~MeV}, &
    m_e \sim  0.486847 {\rm ~MeV}, \\
    m_c \sim  677_{-61}^{+56} {\rm ~MeV}, &
    m_s \sim  93.4_{-13.0}^{+11.8} {\rm ~MeV}, &
    m_\mu \sim  102.75 {\rm ~MeV}, \\
    m_t \sim  181\pm 13 {\rm ~GeV}, &
    m_b \sim  3.00\pm 0.11 {\rm ~GeV}, &
    m_\tau \sim  1.7467 {\rm ~GeV}.
  \end{array}
\end{eqnarray}

\subsection{Mixing angles}
The observed values for the CKM matrix elements are \cite{PDG}
\begin{equation}
  \left|V_{\rm CKM}\right| = \pmatrix{
    0.9745-0.9757  &  0.219-0.224    &  0.002-0.005  \cr
    0.218-0.224    &  0.9736-0.9750  &  0.036-0.046  \cr
    0.004-0.014    &  0.034-0.046    &  0.9989-0.9993   }.
\end{equation}
These data also indicate hierarchical structure:
\begin{equation}
  \theta_{12} \sim \sin \theta_C \equiv \lambda \sim 0.22, \qquad
  \theta_{23} \sim \lambda^2, \qquad
  \theta_{13} \equiv x \sim \lambda^{3-4}.
\end{equation}
It is interesting to note the following relations between the mixing
angles and the relevant mass eigenvalues (at $M_Z$ scale):
\begin{itemize}
\item $\theta_{12} \sim 0.22$
  \begin{equation}
    \sqrt{m_u/m_c} \,=\, 0.051 \sim 0.067, \qquad
    \sqrt{m_d/m_s} \,=\, 0.196 \sim 0.256,
  \end{equation}
\item $\theta_{13} \sim 0.003$
  \begin{equation}
    \sqrt{m_u/m_t} \,=\, 0.003 \sim 0.004, \qquad
    \sqrt{m_d/m_b} \,=\, 0.036 \sim 0.043,
  \end{equation}
\item $\theta_{23} \sim 0.037$
  \begin{equation}
    \sqrt{m_c/m_t} \>=\, 0.056 \sim 0.066, \qquad
    \sqrt{m_s/m_b} \>=\, 0.161 \sim 0.191,
    \label{2-3}
  \end{equation}
\end{itemize}
where the mixing angles may be related to the ratios of the mass
eigenvalues by taking the following down \cite{1-2mixing} and up
\cite{1-3mixing} side mass matrices via the seesaw mechanism:
\begin{equation}
  m_D^{(12)} \,=\, \bordermatrix{
       &    1    &     2     \cr
    1  &    0    &  \lambda  \cr
    2  & \lambda &     1     \cr  }\cdot m_s,
\end{equation}
and
\begin{equation}
  m_U^{(13)} \,=\, \bordermatrix{
       &  1  &  3  \cr
    1  &  0  &  x  \cr
    3  &  x  &  1  \cr  }\cdot m_t.
\end{equation}
However, the 2-3 mixing is too small compared with the mass ratio in
either the up or down sector. We need a more complicated mechanism
which may be the combination of mixing in both up and down
sectors.

\subsection{Fermion masses of third generation in ESSM}
Before proceeding to the analysis of the full fermion masses in the
ESSM we make a quick review of our previous results for the third
generation fermion masses.

The characteristic features of the ESSM are that, due to the ANF
character, the Yukawa couplings approach their infrared fixed points
very rapidly and that the RGE effect of QCD on the quark enhances down
to charged lepton mass ratio by a factor of approximately 
$5-6$\@. This is much more than in the MSSM case, in which this factor
is $\sim 3$\@. One might think that this QCD enhancement will make it
impossible to bring the low-energy bottom-tau mass ratio
$R_{b/\tau}(M_Z)$ down to the experimental value ($1.6 \sim 1.8$) even
with large Yukawa couplings (large $\tan \beta$). \ However, if we
adopt the unification condition of an $SO(10)$ GUT with a
$\overline{126}$-Higgs, the extra enhancement from QCD is actually
welcome, since $R_{b/\tau}$ must be enhanced by a factor of $5-6$ to
reproduce the experimental value of $R_{b/\tau}$,
\begin{equation}
  Y_t(M_{\rm GUT}) \,=\, Y_b(M_{\rm GUT}) \,=\, \frac{1}{3}
    Y_{\tau}(M_{\rm GUT}) \;\;\rightarrow\;\; R_{b/\tau}(M_Z) \,\sim\,
    \frac{5 \sim 6}{3}.
\end{equation}

Another remarkable result is that due to the ANF gauge couplings, the
top and bottom Yukawa couplings are determined almost independently of
their initial values fixed at GUT scale. Indeed these Yukawa couplings
reach to their fixed points, which are physically significant and
provide us with reliable predictions of low-energy parameters. By
using these fixed-point solutions and the experimental value of 
$\alpha_3(1 \mbox{TeV}) \sim \, 0.093$, we get, for
example,\footnote{The tau-lepton Yukawa coupling generally has no
  infrared fixed-point solution since it does not have a strong
  $SU(3)$ interaction. We then treat it as an input parameter and
  determine the value of $\tan \beta$ from the experimental value for
  $m_\tau(M_Z)$.}
\begin{eqnarray}
  &&m_t(M_Z) \,\sim\, 178 \,\mbox{GeV},\quad m_b(M_Z) \,\sim\, 3.2
  \,\mbox{GeV}. \\
  &&\hspace*{2.5cm}( \tan \beta \sim 58 ) \nonumber
\end{eqnarray}
These values are certainly consistent with the experimental
values \cite{PDG}.

\section{Mass texture and IR fixed points}
\setcounter{equation}{0}\setcounter{footnote}{0}

We consider the following extended supersymmetric standard model
with 5 generations, the MSSM (3 generations) + 1 extra vector-like
family ($4 + \bar 4$). The matter content of this model is
\begin{eqnarray}
  && Q_i,\, u_i,\, d_i,\, L_i,\, e_i\,, \quad (i = 1,\cdots,4) \\
  && \bar Q,\, \bar u,\, \bar d,\, \bar L,\, \bar e\,, \qquad (i =
  \bar 4) \\
  && H,\, \bar H,\, \Phi\,.
\end{eqnarray}
In addition to $H$ and $\bar H$, which form a pair of $SU(2)_W$
doublet Higgs fields, we have $\Phi$, a (standard gauge group) singlet
Higgs which yields masses of the extra vector-like family. In this
paper, we consider one $\Phi$, which attains a vacuum expectation
value on the order of the TeV scale \cite{tc}-\cite{invmass}, by
taking suitably soft SUSY breaking terms of $\Phi$\@. In this
situation, the superpotential becomes
\begin{eqnarray}
  W &=& \sum_{i,j=1,\cdots,4} \left({\mbox{\boldmath
    $Y$}}_{\!u_{\,ij}} \epsilon_{ab} Q^a_i u_j \bar H^b +
    {\mbox{\boldmath $Y$}}_{\!d_{\,ij}} \epsilon_{ab} Q^a_i d_j H^b +
    {\mbox{\boldmath $Y$}}_{\!e_{\,ij}} \epsilon_{ab} L^a_i e_j H^b
    \right) \nonumber \\[1mm]
  &&\hspace*{1cm} +\, Y_{\bar u} \bar Q_a \bar u H^a + Y_{\bar d} \bar
  Q_a \bar d \bar H^a + Y_{\bar e} \bar L_a \bar e \bar H^a + Y \Phi^3 
  \nonumber \\[3mm]
  && + \sum_{i=1,\cdots,4} \left(Y_{Q_i} \Phi Q^a_i \bar Q_a + Y_{u_i}
    \Phi u_i \bar u + Y_{d_i} \Phi d_i \bar d + Y_{L_i} \Phi L^a_i
    \bar L_a + Y_{e_i} \Phi e_i \bar e\, \right),\qquad \quad
  \label{spot}
\end{eqnarray}
where the subscripts $i,j$ and $a,b$ are indices of generation and
$SU(2)_W$, respectively, and other indices are trivially
contracted. With this superpotential, after $SU(2)_W \times U(1)_Y$
breaking, the forms of the $5\times 5$ fermion mass matrices can be
written as
\begin{eqnarray}
  m_U &=& \bordermatrix{
           & u_{1R} & \cdots & u_{4R} & u_{\bar 4 R}  \cr
    u_{1L} &        &        &        &               \cr
    \;\;\vdots &    & {\mbox{\boldmath $Y$}}_{\!u_{\,ij}} v_u  & &
           Y_{Q_i} V \cr 
    u_{4L} &        &        &        &               \cr
    u_{\bar 4 L} &  & Y_{u_i} V &  & Y_{\bar u} v_d   \cr  }\,,
  \\
  m_D &=& \bordermatrix{
           & d_{1R} & \cdots & d_{4R} & d_{\bar 4 R}  \cr
    d_{1L} &        &        &        &               \cr
    \;\;\vdots &    & {\mbox{\boldmath $Y$}}_{\!d_{\,ij}} v_d  & &
           Y_{Q_i} V \cr 
    d_{4L} &        &        &        &               \cr
    d_{\bar 4 L} &  & Y_{d_i} V &  & Y_{\bar d} v_u   \cr   }\,,
  \\
  m_E &=& \bordermatrix{
           & e_{1R} & \cdots & e_{4R} & e_{\bar 4 R}  \cr
    e_{1L} &        &        &        &               \cr
    \;\;\vdots &    & {\mbox{\boldmath $Y$}}_{\!e_{\,ij}} v_d  & &
           Y_{L_i} V \cr 
    e_{4L} &        &        &        &               \cr
    e_{\bar 4 L} &  & Y_{e_i} V &  & Y_{\bar e} v_u   \cr   }\,,
\end{eqnarray}
\begin{eqnarray}
  \langle H \rangle = \pmatrix{ v_d \cr 0 }, \quad
  \langle \bar H \rangle = \pmatrix{ 0 \cr v_u}, \quad
  \langle\Phi\rangle = V,
\end{eqnarray}
where $u_{iL}, d_{iL}, e_{iL}, (u_{\bar 4 R})^C, (d_{\bar 4 R})^C$ and 
$(e_{\bar 4 R})^C$ are fermionic components of $SU(2)_W$ doublet
fields, and $(u_{iR})^C, (d_{iR})^C, (e_{iR})^C, u_{\bar 4 L}, d_{\bar
  4 L}$ and $e_{\bar 4 L}$ are those of $SU(2)_W$ singlets.

\subsection{Candidate for texture and IR fixed points}

Before discussing realistic texture, we classify the types of texture
which yield hierarchical masses, referring to their infrared
behavior. Since we know that the usual quark and lepton mass matrices
exhibit generally typical hierarchical structures, we first consider
the dominant part of the matrices including only the 3rd, 4th and
$\bar 4$th generations and then include less important contributions
step-by-step. Hereafter, for simplicity $m\,,\bar m$ and $M$ are used
symbolically to represent ${\mbox{\boldmath $Y$}}_{\!f_{\,ij}}
v_{u,d}\,,Y_{\bar f}\, v_{u,d}$ and $Y_{f_i} V \,(f=u,d,e)$,
respectively, since in our classification only the order of their
masses are important $(m \ll M)$. For the moment, let us restrict
ourselves to the situation in which Yukawa couplings,
${\mbox{\boldmath $Y$}}_{\!f_{\,ij}}$, are symmetric (at the GUT
scale) and only the 4th generation couples to the $\bar 4$th
generation forming an invariant mass term $M$. Analyses for general
situations are performed in Appendix A.

First let us consider the dominant matrices (for the 3rd, 4th and
$\bar 4$th generations). For $m \ll M$, after diagonalization at low
energy, two of the three eigenvalues are on the order of $M$, and one
eigenvalue $m_{33}$ is small compared with $M$. In order to get
non-zero $m_{33}$, there are three candidates for the textures
classified by their determinants:
\begin{itemize}
\item case 1
  \begin{eqnarray}
    \bordermatrix{
             &  ~3  &  4  &  \bar 4  \cr
      3  ~   &  ~m  &     &   0      \cr
      4      &      &     &   M      \cr
      \bar 4 &  ~0  &  M  &          \cr
      },
    \qquad \det~_{\hspace*{-1.5mm}3\times 3} \sim M^2 m
    \label{matrix:1}
  \end{eqnarray}
  \begin{eqnarray}
    \hspace*{-1cm} m_{33} \sim m.
  \end{eqnarray}
\item case 2
  \begin{eqnarray}
    \bordermatrix{
             &  ~3  &  ~4  &  \bar 4  \cr
      3 ~    &  ~0  &  ~m  &   0      \cr
      4      &  ~m  &      &   M      \cr
      \bar 4 &  ~0  &  ~M  &  \bar m  \cr
      },
    \qquad \det~_{\hspace*{-1.5mm}3\times 3} \sim m^2 \bar m
    \label{matrix:2}
  \end{eqnarray}
  \begin{eqnarray}
    \hspace*{-1cm} m_{33} \sim \left( \frac{m \bar m}{M^2} \right) m.
  \end{eqnarray}
\item case 3
  \begin{eqnarray}
    \bordermatrix{
             &  ~3  &  ~4  &  \bar 4  \cr
      3 ~    &  ~0  &  ~m  &   0      \cr
      4      &  ~m  &  ~m  &   M      \cr
      \bar 4 &  ~0  &  ~M  &   0      \cr
      },
    \qquad \det~_{\hspace*{-1.5mm}3\times 3} \sim 0
    \label{matrix:3}
  \end{eqnarray}
  \begin{eqnarray}
    m_{33} \sim \mbox{(radiatively induced)}.
  \end{eqnarray}
\end{itemize}
Here and hereafter, if an element is 0 it is implied that it should be 
exactly zero at the GUT scale. A blank entry corresponds to the case
in which it can be either zero or nonzero. For the case 3, the
determinant of this texture is zero. However, this type of matrix
induces an appreciable non-zero $Y_{33}$ element via the
renormalization group, and the resultant low-energy determinant (and
eigenvalue $m_{33}$) cannot be neglected.

Next, we include the second generation and consider $4\times 4$
matrices. In order to obtain hierarchical mass eigenvalues after
diagonalization, $m_{22} \ll m_{33} \ll M$, it is needed to realize
the hierarchical determinants $\det~_{\hspace*{-1.5mm}4\times
  4}/\det~_{\hspace*{-1.5mm}3\times 3} \ll m_{33}$\@. Then the
resultant $4\times 4$ textures which realize this hierarchical
structure are found to be the followings for each case:
\begin{itemize}
\item case 1

  There are two distinct types of texture.
  \begin{itemize}
  \item[$\circ$] type A
    \begin{eqnarray}
      \bordermatrix{
               &  2  &  3   &  4  &  \bar 4 \cr
        2 ~    &  0  &  0   &  m  &   0     \cr
        3      &  0  &  m   &     &   0     \cr
        4      &  m  &      &     &   M     \cr
        \bar 4 &  0  &  0   &  M  &  \bar m \cr
        },
      \qquad \det~_{\hspace*{-1.5mm}4\times 4} \sim m^3 \bar m
      \label{matrix:A}
    \end{eqnarray}
    \begin{eqnarray}
      \hspace*{-1cm} m_{22} \sim \left( \frac{m \bar m}{M^2} \right)
      m.
      \label{m22A}
    \end{eqnarray}
  \item[$\circ$] type B
    \begin{eqnarray}
      \bordermatrix{
               &  2  &  3   &  4  &  \bar 4 \cr
        2 ~    &  0  &  0   &  m  &   0     \cr
        3      &  0  &  m   &     &   0     \cr
        4      &  m  &      &  m  &   M     \cr
        \bar 4 &  0  &  0   &  M  &   0     \cr
        },
      \qquad \det~_{\hspace*{-1.5mm}4\times 4} \sim 0
      \label{matrix:B}
    \end{eqnarray}
    \begin{eqnarray}
      m_{22} \sim \mbox{(radiatively induced)}.
    \end{eqnarray}
  \end{itemize}
  Note that in these textures the 3rd generation is almost decoupled
  from the other ones. Contrastingly, these textures are of the same
  form as those of cases 2 and 3 ((\ref{matrix:2}) and
  (\ref{matrix:3})) which are used for the 2nd, 4th and $\bar 4$th
  generations to obtain non-zero $m_{22}$\@. Therefore, these small 
  eigenvalues $m_{22}$ are strongly affected by the existence of the
  4th and $\bar 4$th generations.
\item case 2, 3

  We found that there is essentially no candidate in these cases.

  One might think that the texture
  \begin{eqnarray}
    \bordermatrix{
             &  2  &  3   &  4  &  \bar 4  \cr
      2 ~    &  0  &  0   &  m  &   0      \cr
      3      &  0  &  0   &  m  &   0      \cr
      4      &  m  &  m   &  m  &   M      \cr
      \bar 4 &  0  &  0   &  M  &           \cr
      },
    \qquad \det~_{\hspace*{-1.5mm}4\times 4} \sim 0,
  \end{eqnarray}
  reproduces a non-zero eigenvalue $m_{22}$ radiatively. However, in
  this texture the second and third generations have the same
  structure. Therefore this matrix has at most rank 3 and yields an
  almost zero eigenvalue for $m_{22}$ even when all the radiatively
  induced Yukawa couplings are included.

  There are two more candidates for texture having hierarchical
  mass structure:
  \begin{eqnarray}
    \bordermatrix{
             &  2  &  3   &  4  &  \bar 4  \cr
      2 ~    &  m  &  0   &     &   0      \cr
      3      &  0  &  0   &  m  &   0      \cr
      4      &     &  m   &     &   M      \cr
      \bar 4 &  0  &  0   &  M  &  \bar m  \cr
      },
    &&\qquad \det~_{\hspace*{-1.5mm}4\times 4} \sim m^3 \bar m,
    \label{matrix:X1}  \\[3mm]
    \bordermatrix{
             &  2  &  3   &  4  &  \bar 4 \cr
      2 ~    &  m  &  0   &     &   0     \cr
      3      &  0  &  0   &  m  &   0     \cr
      4      &     &  m   &  m  &   M     \cr
      \bar 4 &  0  &  0   &  M  &   0     \cr
      },
    &&\qquad \det~_{\hspace*{-1.5mm}4\times 4} \sim 0.
    \label{matrix:X2}
  \end{eqnarray}
  However, (\ref{matrix:X1}) and (\ref{matrix:X2}) coincide with
  (\ref{matrix:A}) and (\ref{matrix:B}) by changing the label of the
  generation ($2\leftrightarrow 3$). We do not consider these types of
  texture.\footnote{Strictly speaking, we cannot exchange the
    generation label ($2\leftrightarrow 3$) when we use one of the
    textures (\ref{matrix:A}) and (\ref{matrix:B}) for the up-quark
    sector and one of (\ref{matrix:X1}) and (\ref{matrix:X2}) for the
    down-quark, and vice versa, for example. This case yields a large
    mixing angle between generations, which is experimentally
    excluded.}
\end{itemize}

In conclusion, we have two candidates, (\ref{matrix:A}) and
(\ref{matrix:B}), which may reproduce hierarchical mass structure
between the second and third generations of the up and down sectors.

\begin{figure}[htbp]
  \parbox{6.6cm}{
    \epsfxsize=5.6cm \ \epsfbox{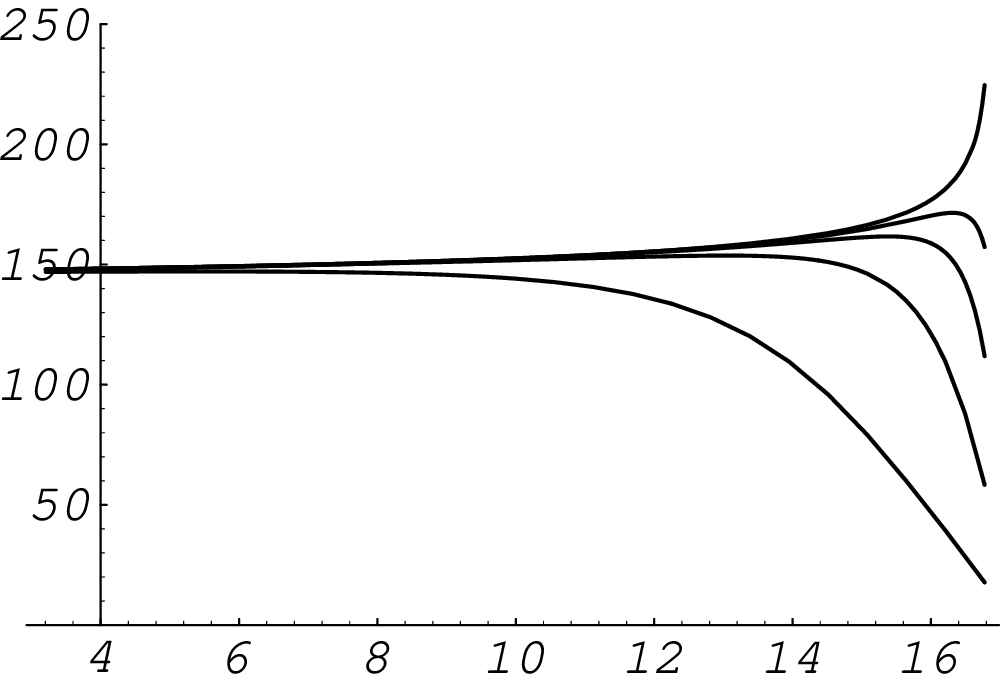}
    \put(-160,121){\footnotesize $m_{33}/g_3$}
    \put(5,95){\footnotesize $Y_i=5$}
    \put(7,67){\footnotesize $3$}
    \put(7,50){\footnotesize $2$}
    \put(7,33){\footnotesize $1$}
    \put(7,20){\footnotesize $0.5$}
    \put(7,8){\footnotesize $\log_{10} \mu$}
    \vspace*{-2mm}
    \caption{Typical behavior of $m_{33}$ in the type A texture.}
    \label{fig:33A}
    }
  \hspace*{10mm}
  \parbox{6.6cm}{
    \epsfxsize=5.6cm \ \epsfbox{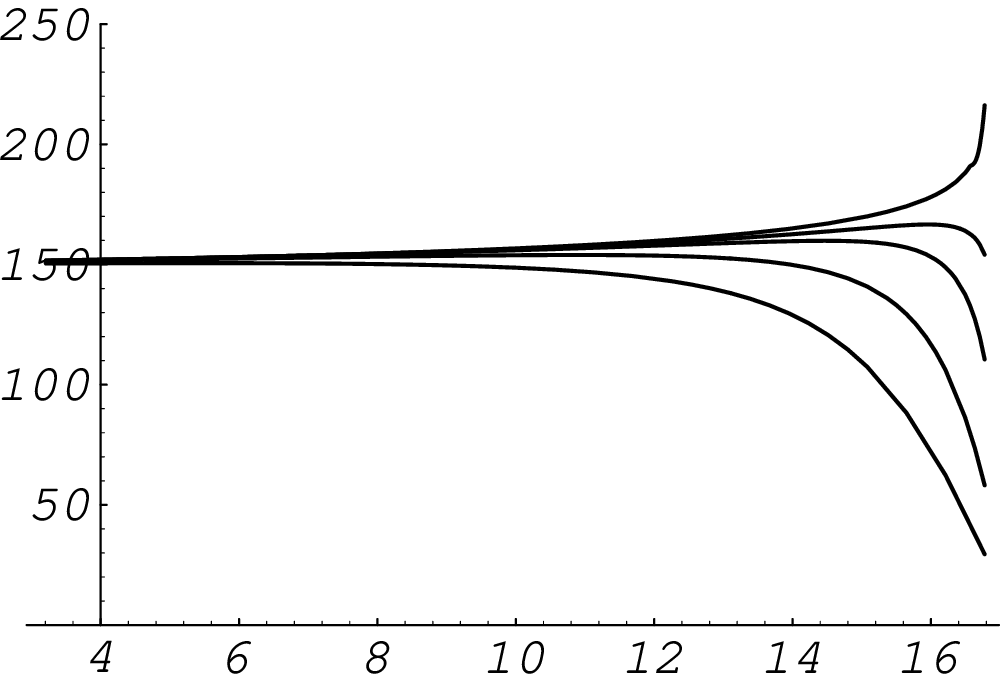}
    \put(-160,121){\footnotesize $m_{33}/g_3$}
    \put(5,93){\footnotesize $Y_i=5$}
    \put(7,67){\footnotesize $3$}
    \put(7,49){\footnotesize $2$}
    \put(7,32){\footnotesize $1$}
    \put(7,19){\footnotesize $0.5$}
    \put(7,8){\footnotesize $\log_{10} \mu$}
    \vspace*{-2mm}
    \caption{Typical behavior of $m_{33}$ in the type B texture.}
    \label{fig:33B}
    } 
\end{figure}

\begin{figure}[htbp]
  \parbox{6.6cm}{
    \epsfxsize=5.6cm \ \epsfbox{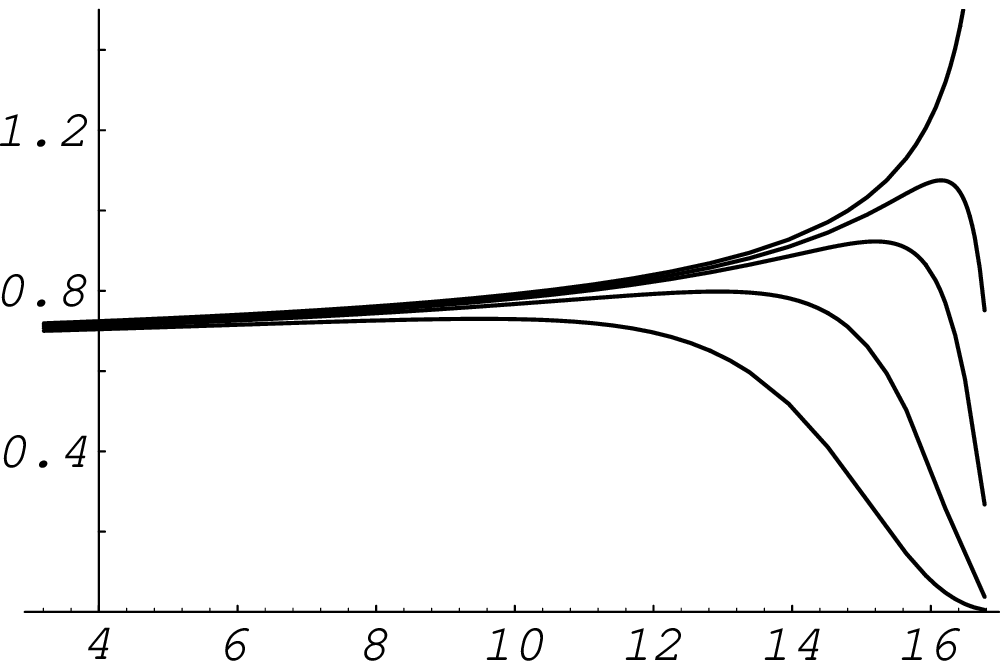}
    \put(-160,119){\footnotesize $m_{22}/g_3$}
    \put(3,105){\footnotesize $Y_i=5$}
    \put(5,57){\footnotesize $3$}
    \put(5,30){\footnotesize $2$}
    \put(5,17){\footnotesize $1$}
    \put(5,8){\footnotesize $0.5$}
    \caption{Typical behavior of $m_{22}$ in the type A texture.}
    \label{fig:22A}
    }
  \hspace*{10mm}
  \parbox{6.6cm}{
    \epsfxsize=5.6cm \ \epsfbox{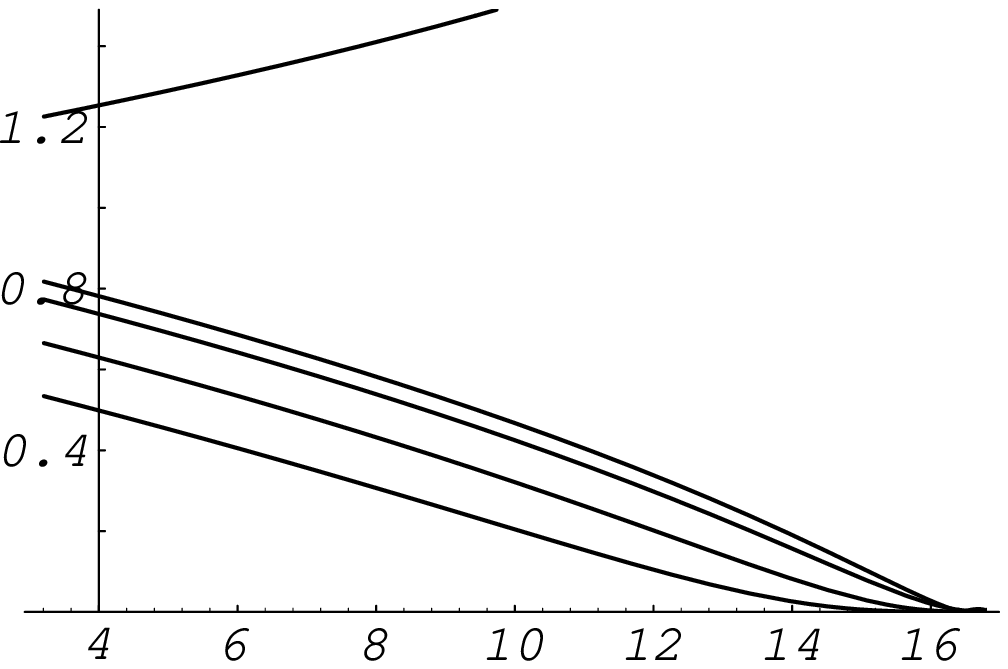}
    \put(-160,119){\footnotesize $m_{22}/g_3$}
    \put(-72,107){\footnotesize $Y_i=5$}
    \put(-5,30){\footnotesize $Y_i=3$}
    \put(-4.5,20){\footnotesize $\hspace*{5.2ex}2$}
    \put(-4.5,10){\footnotesize $\hspace*{5.2ex}1$}
    \put(-8,0){\footnotesize $\hspace*{5.2ex}0.5$}
    \caption{Typical behavior of $m_{22}$ in the type B texture.}
    \label{fig:22B}
    } \vspace*{3mm}

{\small
(In the above figures, we set $M_{\rm GUT} = 7\times 10^{16}$ GeV,
$\alpha_{\rm GUT} = 1.0$, $M = 1$ TeV, $\tan \beta = 5$, and all the
non-zero Yukawa couplings at $M_{\rm GUT}$ are taken to have the same
value $Y_i$.)}
\end{figure}

In addition to the difference between the mass eigenvalues for types A
and B, another remarkable difference exists between these two types of
texture. As we have already mentioned, it is the infrared behavior of
Yukawa couplings that is characteristic to this ANF model. As long as
we restrict ourselves to the third generation, two textures have the
same infrared structures (Figs.\ \ref{fig:33A} and \ref{fig:33B}): the
Yukawa couplings $Y_{33}$ sit almost on their infrared fixed points at
low energy in both cases. However, the infrared behavior of the
eigenvalues of the second generation is quite different. For the type
A texture, the resultant eigenvalue $m_{22}$ (see (\ref{m22A})) is
obtained from the tree-level Yukawa couplings (and VEVs whose orders
are assumed), and therefore it can be regarded as an infrared fixed
point value (Fig.\ \ref{fig:22A}). On the other hand, for the type B
texture, the eigenvalue $m_{22}$ is induced radiatively by the
renormalization procedure and does not reach its theoretical infrared
fixed point value at low energy (Fig.\ \ref{fig:22B}).

\subsection{Texture for mass matrix}

\begin{figure}[htbp]
  \parbox{6.6cm}{
    \epsfxsize=5.7cm \ \epsfbox{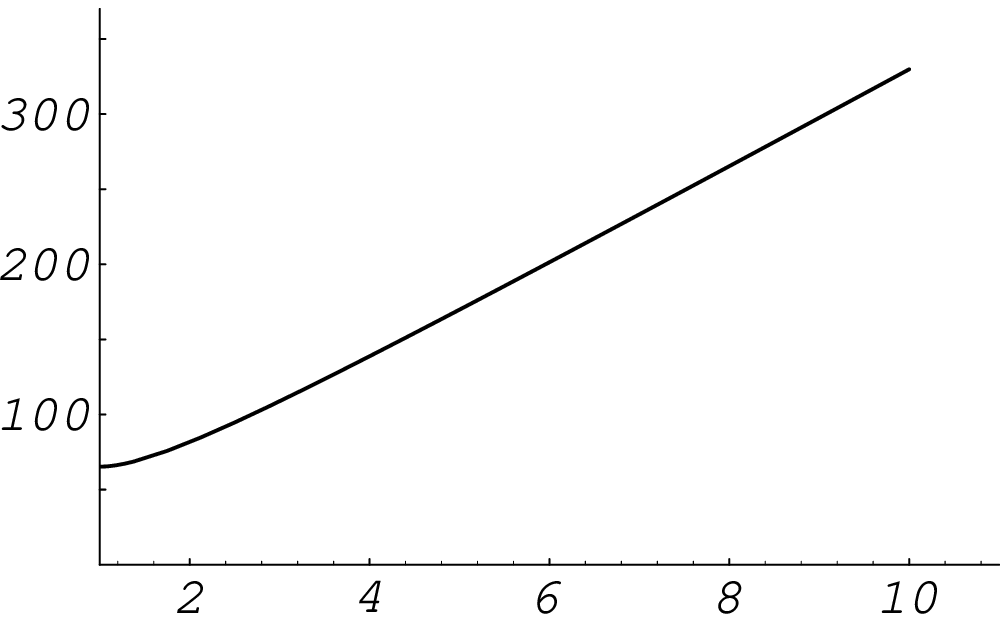}
    \put(6,10){\footnotesize $\tan \beta$}
    \put(-164,117){\footnotesize
      $\displaystyle{\left(\frac{m_{33}}{m_{22}}\right) \simeq
      \left(\frac{m \bar m}{M^2} \right)^{-1}}$}
    \caption{Typical behavior of the hierarchical factor in the type A
    texture ($\alpha_{\rm GUT} = 1$, $M=1$ TeV).}
    \label{fig:sup}
    }
  \hspace*{10mm}
  \parbox{6.6cm}{
    \epsfxsize=5.7cm \ \epsfbox{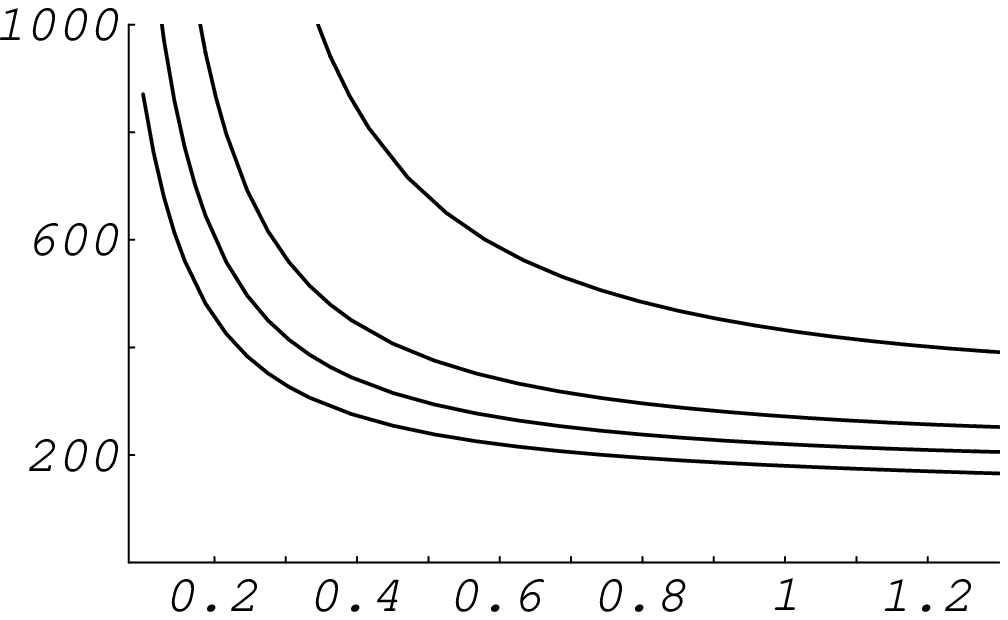}
    \put(6,8){\small $\alpha_{\rm GUT}$}
    \put(-12.5,55){\footnotesize $Y_i = 0.2$}
    \put(10,36){\footnotesize $0.5$}
    \put(13,26.5){\footnotesize $1$}
    \put(13,18){\footnotesize $3$}
    \put(-156,118){\footnotesize
      $\displaystyle{\left(\frac{m_{33}}{m_{22}}\right)}$}
    \caption{Typical behavior of the hierarchical factor in the type B
  texture ($\tan \beta =5$, $M=1$ TeV).}
    \label{fig:ratio}
    }
\end{figure}

Next, we argue which texture can be used for the quark and lepton
sectors. From the above discussions we have learned that the type A
texture provides us with a mechanism in which we can make a full use
of infrared fixed points. We see from figure \ref{fig:sup} that the
typical value of the hierarchical factor $\frac{m \bar m}{M^2}$ (see
(\ref{m22A})) is generally $1/100$ or less, and that this factor
becomes smaller for larger $\tan \beta$.
It is far smaller than the observed strange to bottom mass ratio
$(\sim 1/30)$. Therefore we do not use the type A texture but adopt
the type B one for the down-quark and charged lepton sectors. On the
other hand, the hierarchical factor in the up-quark sector, charm to
top mass ratio ($\sim 1/250$), is much smaller than the ratio of
bottom and strange. We can adopt the type A mass texture for the
up-quark sector. Note that in the type A texture, the masses of heavy
quarks (charm as well as top) are given by the infrared fixed point
values for Yukawa couplings, which are almost insensitive to their
initial values at $M_{\rm GUT}$.

Once we fix the texture for the up and down/lepton sectors, we can
estimate the quark and lepton masses at low-energy scale. However,
there arise two difficulties in reproducing hierarchical structures 
both in up and down sectors. First, the top and bottom Yukawa
couplings reach their infrared fixed point values, which requires a
large $\tan \beta$ scenario. This large $\tan \beta$ scenario makes
the hierarchical factor in the up sector much smaller than the actual
value of the ratio of top and charm (see Fig.\
\ref{fig:sup})\@. Second, the ratio of the eigenvalues,
$m_{33}/m_{22}$, at low energy is found to be at least 100 in the type
B texture (Fig.\ \ref{fig:ratio}).

Some improvement must be made to overcome the above mismatches. It is
found that we can remove both difficulties by introducing only one
parameter, $\epsilon$, and attaching it to bottom/tau Yukawa
couplings. Even when $\epsilon$ is introduced, the type A texture
cannot be used for the down/lepton sectors, because their non-zero
$Y_{\bar 4 \bar 4}$ elements contribute to the beta-function of the
top Yukawa coupling and make its infrared fixed point value far
smaller than the experimental one. Then, we finally find that the
following mass texture for quark and lepton at the GUT scale can
reproduce the low-energy experimental values of the fermion masses:
\begin{eqnarray}
  m_U &=& \bordermatrix{
           &  2  &  3   &  4  &  \bar 4 \cr
    2 ~    &  0  &  0   &  m  &   0     \cr
    3      &  0  &  m   &     &   0     \cr
    4      &  m  &      &     &   M     \cr
    \bar 4 &  0  &  0   &  M  &  \bar m \cr
    }\,,
  \label{matrix:mU}
  \\
  m_D &=& \bordermatrix{
           &  2  &  3   &  4  &  \bar 4 \cr
    2 ~    &  0  &  0   &  m  &   0     \cr
    3      &  0  & \epsilon m & 0 & 0   \cr
    4      &  m  &  0   &  m  &   M     \cr
    \bar 4 &  0  &  0   &  M  &   0     \cr
    }\,,
  \\
  m_E &=& \bordermatrix{
           &  2  &  3   &  4  &  \bar 4  \cr
    2 ~    &  0  &  0   &  3 m  &   0    \cr
    3      &  0  &  3 \epsilon m & 0 & 0 \cr
    4      &  3 m & 0   &  3 m  &   M    \cr
    \bar 4 &  0  &  0   &  M  &   0      \cr
    }\,.
\end{eqnarray}
Here we have used the Higgs field of the $\overline{126}$
($\overline{45}$) representation of $SO(10)$ ($SU(5)$) so that the
boundary conditions in the down/lepton sector may correctly reproduce
the observed bottom/tau (and strange/mu) ratio, as noted in the
previous section. In this texture, all the quark Yukawa couplings
except for ${\mbox{\boldmath $Y$}}_{\!d_{\,33}}$ converge to their
infrared fixed points independently of their initial values at the GUT
scale. Within this approximation, this texture leads to the low-energy
prediction of fermion masses:\footnote{For simplicity, we set the
  blanks in $m_U$ (\ref{matrix:mU}) to be zeros.}
\begin{eqnarray}
  \begin{array}{lcl}
    m_t \;\sim\, 180 {\rm ~GeV}, && m_c \;\sim\, 1.0 {\rm ~GeV},\\
    m_b \>\sim\, 3.1 {\rm ~GeV}, && m_s \>\sim\; 0.081 {\rm ~GeV},\\
    m_\tau \,\sim\, 1.75 {\rm ~GeV}, && m_\mu \,\sim\, 0.103 {\rm
      ~GeV},
  \end{array}
  \quad (\mbox{at} \;M_Z)
\end{eqnarray}
for the input values
\begin{eqnarray}
  \begin{array}{lll}
    M_{\rm GUT} \,\sim\, 5.3 \times 10^{16} ~{\rm GeV}, ~&
    \alpha_{\rm GUT} \sim 0.3\,, ~&
    \epsilon \,\sim\, 0.2\,, \\[1mm]
    M_{\rm SUSY} \sim 1 ~{\rm TeV}, &
    \tan \,\beta\, \sim 20\,, &
    V \sim 3 ~{\rm TeV}.
  \end{array}
  \label{input}
\end{eqnarray}
These results are surely in good agreement with the present
experimental values. For the above input values, both $M_{\rm GUT}$
and $\alpha_{\rm GUT}$ are larger than those of the usual MSSM because
of the asymptotically non-free character of this ESSM\@. It is also
noted that $\epsilon$ is on the order of the Cabibbo angle. This fact
may be naturally reproduced with the anomalous $U(1)$ symmetry
\cite{anomalous}, which may be helpful in considering hierarchies
between the first and other generations.

\subsection{CKM mixing angles}

Encouraged by the successful predictions of the texture in the
previous section, we finally consider full mass matrices including
quark mixing angles. This may be done by introducing hierarchically
very small Yukawa couplings. As for these very small couplings, we
cannot apply the IRFP approach to texture forms in a similar way to
the previous section. However, by considering the extension of the
above uniquely obtained texture, we can almost fix the Yukawa
interactions between all five generations phenomenologically.

As noted in section 2, the 1-3 quark mixing is described well by the
1-3 mixing in the up sector and the 1-2 quark mixing by the 1-2 mixing
in the down sector. Therefore the eigenvalue $m_{11}$ for up- (down-)
quark sector can be automatically reproduced by introducing a small
mixing ${\mbox{\boldmath $Y$}}_{\!u_{\>\!13}} \,({\mbox{\boldmath
    $Y$}}_{\!d_{\,12}})$ into our texture. A problem arises in the 2-3
quark mixing for which the usual seesaw-type relation between mixing
angle and eigenvalues is not successful. Fortunately, however, in our
extended model, the mass eigenvalues for the second generation can be
properly reproduced by mixing with the extra vector-like
generations. Therefore the 2-3 mixing parameter can be treated
independently of the 2-2 mass eigenvalue as long as it does not affect
$m_{22}$ very strongly. We can thus adopt the following $3\times 3$
matrices for the ordinary generations in which the CKM mixing angles
may correctly reproduced.
\begin{eqnarray}
  (m_U)_{3\times 3} &\,=\,& \bordermatrix{
        & ~1 &  2  & 3~          \cr
    1 ~ &    &     & \epsilon^l~ \cr
    2   &    &     &             \cr
    3   & ~\epsilon^l & & 1~ }\cdot m_t\,,
  \\
  (m_D)_{3\times 3} &\,=\,& \bordermatrix{
        &   1   &   2    &   3        \cr
    1 ~ &   & \epsilon^m &            \cr
    2   & \epsilon^m  &  & \epsilon^n \cr
    3   &   & \epsilon^n &     1      }\cdot m_b\,.
\end{eqnarray}
The down-quark (and charged lepton) sector is just the Fritzsch type
of texture \cite{Fritzsch}, in which the mass of the second generation
induced by the seesaw mechanism is much smaller than ``tree-level''
one (see (\ref{2-3})), which now comes from mixing with the extra
generations. At this point, note that $\epsilon$ in the texture in the
previous section happens to have just the same value as the Cabibbo
mixing angle for explaining all hierarchies in the second and third
generations of quarks and leptons. Therefore it is not necessary to
introduce any small mixing parameter other than $\epsilon$.
\begin{eqnarray}
  \begin{array}{c|c} \hline \hline
    m & |V_{us}| \\ \hline
    2 & 0.6  \\
    3 & 0.27 \\
    4 & 0.06 \\ \hline
  \end{array}
  \qquad
  \begin{array}{c|c} \hline \hline
    n & |V_{cb}| \\ \hline
    1 & 0.17  \\
    2 & 0.035 \\
    3 & 0.007 \\ \hline
  \end{array}
  \qquad
  \begin{array}{c|c|c} \hline \hline
    l & |V_{ub}| & m_u \\ \hline
    3 & 0.008  & \sim 32  \;{\rm MeV} \\
    4 & 0.002  & \sim 1.3 \;{\rm MeV} \\
    5 & 0.0003 & \sim 0.5 \;{\rm MeV} \\ \hline
  \end{array}
\end{eqnarray}
\begin{eqnarray}
  \begin{array}{c|c|c||c|c|c||c|c|c} \hline \hline
    m & n & |V_{ub}| & m & n & |V_{ub}| & m & n & |V_{ub}| \\ \hline
    2 & 1 & 0.044 & 3 & 1 & 0.01   & 4 & 1 & 0.002  \\
    2 & 2 & 0.009 & 3 & 2 & 0.002  & 4 & 2 & 0.0004 \\
    2 & 3 & 0.002 & 3 & 3 & 0.0004 & 4 & 3 & 0.0001 \\ \hline
  \end{array}
\end{eqnarray}
After all, there is a reasonable $5\times 5$ GUT-scale texture which
explains the experimental values of the CKM mixing angles, and we can
see that this texture is actually almost the only possibility left in
this situation.
\begin{eqnarray}
  m_U &\simeq& \bordermatrix{
           &   1   &   2   &   3   &   4   &   \bar 4 \cr
    1 ~~   &   0   &   0   & \epsilon^4 m &  0  &  0  \cr
    2      &   0   &   0   &   0   &   m   &   0      \cr
    3      & \epsilon^4 m &  0  &  m  &  0 &   0      \cr
    4      &   0   &   m   &   0   &   0   &   M      \cr
    \bar 4 &   0   &   0   &   0   &   M   &  \bar m  \cr
    }\,,
  \\
  m_D &\simeq&
  \bordermatrix{
           &   1   &   2   &   3   &   4   &   \bar 4    \cr
    1 ~~   &   0   & \epsilon^4 m  &  0  &  0  &  0      \cr
    2      & \epsilon^4 m &  0  & \epsilon^3 m & m &  0  \cr
    3      &   0   & \epsilon^3 m & \epsilon m & 0 &  0  \cr
    4      &   0   &   m   &   0   &   m   &    M        \cr
    \bar 4 &   0   &   0   &   0   &   M   &    0        \cr
    }\,,
  \\
  m_E &\simeq& \bordermatrix{
           &   1   &   2   &   3   &   4   &   \bar 4       \cr
    1 ~~   &   0   & 3\epsilon^4 m &  0  &  0  &  0         \cr
    2      & 3\epsilon^4 m &  0  & 3\epsilon^3 m & 3 m &  0 \cr
    3      &  0    & 3\epsilon^3 m & 3\epsilon m &  0  &  0 \cr
    4      &  0    &   3m  &   0   &   3m  &   M            \cr
    \bar 4 &  0    &   0   &   0   &   M   &   0            \cr
    }\,,
\end{eqnarray}
This texture reproduces the low-energy predictions at $M_Z$ scale:
\begin{eqnarray}
  \begin{array}{lclcl}
    m_u \,\sim\, 2.9 {\rm ~MeV}, &&
    m_d \,\sim\, 4.3 {\rm ~MeV}, &&
    m_e \,\sim\, 0.6 {\rm ~MeV}, \\
    m_c \;\sim\, 1.0 {\rm ~GeV}, &&
    m_s \,\sim\, 0.089{\rm ~GeV}, &&
    m_\mu \,\sim\, 0.104 {\rm ~GeV}, \\
    m_t \;\sim\, 180 {\rm ~GeV}, &&
    m_b \>\sim\, 3.1 {\rm ~GeV}, &&
    m_\tau \,\sim\, 1.75 {\rm ~GeV},
  \end{array}
\end{eqnarray}
\begin{eqnarray}
  \left| V_{\rm CKM} \right| \simeq \pmatrix{
    0.974  &  0.228  &  0.0037 \cr
    0.228  &  0.973  &  0.039 \cr
    0.005  &  0.039  &  0.999 \cr
    }.
\end{eqnarray}
These values are also nearly consistent with the experimental
  ones.\footnote{This $V_{\rm CKM}$ is extracted from the original
  (unitary) $5\times 5$ CKM matrix, and the other matrix elements are
  suppressed by large $M$\@. The unitarity of $V_{\rm CKM}$ is
  realized up to $10^{-4}$\@ for $M\sim O({\rm TeV})$.}

\section{Summary and discussion}

We have investigated the fermion mass matrix structure at the GUT
scale in an asymptotically non-free model with a pair of extra
generations. The characteristic feature of this model is that the
couplings converge to their infrared fixed points very quickly. By
making full use of the IR behavior of the couplings, we determined the
fermion mass matrices at the GUT scale almost uniquely. We have found
the following: i) We can understand the charm quark mass as well as
the top in terms of their infrared fixed point values. It is
interesting that the hierarchical factor of the top and charm ratio 
comes from the existence of the 4th and $\bar 4$th generations at the
TeV scale. Also we should like to note that the Yukawa couplings of
$Y_{24}(Y_{42})$ reach their infrared fixed points with considerable
strength. This indicates that the second generation is strongly
coupled with the extra generations\@. ii) Though the masses of the
other lighter quarks are not related to the infrared structure for the
ANF character, we can determine their mass texture almost uniquely by
introducing only one small parameter. It is interesting that this
small parameter happens to be equal to the Cabibbo mixing angle. In
the down-quark sector the resultant strange-quark mass eigenvalue is
suppressed by the existence of the extra generations, as in the
up-quark sector, in spite of the appreciable large induced Yukawa
coupling $Y_{22}$\@. iii) As for the lepton masses, they are
reproduced quite successfully by assuming that the relevant Higgs
fields belong to $\overline{126}$ representation of $SO(10)$\@. This
is in remarkable contrast to the case of the MSSM in which, as seen
from the Georgi-Jarlskog type of texture \cite{GY}, one has to assume
that the relevant Higgs field must be the mixture of 10 and 126
representations.

In the MSSM case, there are many works containing phenomenological
analyses on the fermion mass structure. It is known that, for example,
the Fritzsch- and Georgi-Jarlskog-type textures provide us with
important hints and standard basis for finding realistic models. Until
now, in the ESSM, we have not established a standard texture which
reproduces the phenomenological fermion masses well, and the aim of
this paper is to establish the form of the possible texture in this
model. Then, we would also like to emphasize the importance of the
IRFP structure. We think that, particularly in asymptotically non-free
theories, the infrared fixed point approach can be one of the
attractive selection rule for the texture form, as well as the other
methods, e.g.\ horizontal symmetry. We could surely assign some
quantum numbers of $U(1)$ symmetry \cite{anomalous} to each fermion
and relevant Higgs particle to reproduce our texture (including
$\epsilon$ parameter), but our aim here is to establish the possible
form of texture first.

It is essential for us to understand the heavier fermion masses as
their IR fixed point values that not only the SUSY breaking scale but
also the invariant masses of the extra generations are on the order of
the TeV scale. This fact implies that when SUSY is discovered, the
extra generations may be also found. Using muon colliders \cite{muon}
in particular, the extra generations may be explored easily, since in
our model the second generation couples strongly to the extra
generations.

\vspace*{5mm}
\subsection*{Acknowledgements}

We would like to thank T.\ Kugo and N.\ Maekawa for many helpful
discussions and valuable comments. M.\ B. is supported in part by the
Grant-in Aid for Scientific Research No. 09640375.

\vspace*{1cm}
\renewcommand{\thesubsection}{Appendix \Alph{subsection}}
\renewcommand{\thesubsubsection}{\arabic{subsubsection}}
\renewcommand{\theequation}{\Alph{subsection}.\arabic{equation}}
\setcounter{section}{0}
\setcounter{equation}{0}\setcounter{footnote}{0}

\subsection{More complex cases}

In this appendix, we briefly present analyses for two more complex
cases. In one case, more than two generations couple to the $\bar 4$th
generation via invariant mass terms and in the other case the form of
the texture is non-symmetric.

\subsubsection{More than two $M$}

Let us make the classification of hierarchical textures, following the
analysis in section 3. First, we consider $3\times 3$ texture. In
addition to the three cases discussed in section 3
((\ref{matrix:1}),$\,$(\ref{matrix:2}),$\,$(\ref{matrix:3})), there
are two types of texture which provide a non-zero eigenvalue
$m_{33}\,$:
\begin{itemize}
\item case 4
  \begin{eqnarray}
    \bordermatrix{
             &  ~3  &  4  &  \bar 4  \cr
      3  ~   &   0  &  m  &   M'     \cr
      4      &   m  &  0  &   M      \cr
      \bar 4 &   M' &  M  &   0      \cr
      },
    \qquad \det~_{\hspace*{-1.5mm}3\times 3} \sim M M' m
    \label{matrix:4}
  \end{eqnarray}
  \begin{eqnarray}
    \hspace*{-1cm} m_{33} \sim \left(\frac{M'}{M}\right) m,
  \end{eqnarray}
\item case 5
  \begin{eqnarray}
    \bordermatrix{
             &  ~3  &  ~4  &  \bar 4  \cr
      3 ~    &   0  &   0  &   M'     \cr
      4      &   0  &   m  &   M      \cr
      \bar 4 &   M' &  ~M  &          \cr
      },
    \qquad \det~_{\hspace*{-1.5mm}3\times 3} \sim M'^2 m
    \label{matrix:5}
  \end{eqnarray}
  \begin{eqnarray}
    \hspace*{-1cm} m_{33} \sim \left(\frac{M'}{M}\right)^2 m.
  \end{eqnarray}
\end{itemize}
{}From the above two textures, we get three types of texture which
produce the hierarchical mass eigenvalues $m_{22} \ll m_{33} \ll M,
M'$ at low energy without a small parameter $\epsilon$, in addition to 
the type A and B textures obtained in section 3.
\begin{itemize}
\item case 4
  \begin{itemize}
  \item[$\circ$] type C
    \begin{eqnarray}
      \bordermatrix{
               &  2  &  3   &  4  &  \bar 4 \cr
        2 ~    &  0  &  m   &  m  &   M'    \cr
        3      &  m  &      &  m  &   0     \cr
        4      &  m  &  m   &     &   M     \cr
        \bar 4 &  M' &  0   &  M  &         \cr
        }\,,
      \label{matrix:C}
    \end{eqnarray}
  \end{itemize}
\item case 5
  \begin{itemize}
  \item[$\circ$] type D
    \begin{eqnarray}
      \bordermatrix{
               &  2  &  3   &  4  &  \bar 4 \cr
        2 ~    &  0  &  m   &  0  &   M'    \cr
        3      &  m  &  m   &  0  &   0     \cr
        4      &  0  &  0   &  m  &   M     \cr
        \bar 4 &  M' &  0   &  M  &         \cr
        }\,,
      \label{matrix:D}
    \end{eqnarray}
  \item[$\circ$] type E
    \begin{eqnarray}
      \bordermatrix{
               &  2  &  3   &  4  &  \bar 4 \cr
        2 ~    &     &  m   &     &   M'    \cr
        3      &  m  &  0   &  m  &   0     \cr
        4      &     &  m   &  m  &   M     \cr
        \bar 4 &  M' &  0   &  M  &         \cr
        }\,.
      \label{matrix:E}
    \end{eqnarray}
  \end{itemize}
\end{itemize}

Next we consider a realistic mass texture of for quarks and
leptons. One can easily see that all the above types of textures have
hierarchical factors of order 1/100 or less and no IRFP structure for
the 2nd generation, unlike the type A texture. Therefore we must also
introduce a (small) parameter $\epsilon$ to obtain the hierarchy in
the down-quark and/or lepton sector. On the other hand, we can get the
hierarchy in the up-quark sector from any one of five textures (type
A--E)\@. As mentioned in section 3, we adopt the type A texture for
the up-quark sector by making active use of the infrared fixed-point
structure which is a characteristic feature of this ANF model. Note
that due to the $SO(10)$ GUT-like relations for Yukawa couplings to a
singlet Higgs, if we adopt the textures of type C--E for the down and
lepton sectors we have to add $M'$ to the up-quark texture (the type A
texture)\@. These quantities $M'$ change the determinant and thus
spoil the hierarchical structure of the type A texture. Therefore we
also must include the small parameter $\epsilon$ in the invariant mass
term $M'$\@. Taking into account all issues discussed to this point,
we searched the realistic textures of quark and lepton at the GUT
scale and found no candidate to reproduce the present experimental
data of hierarchical mass eigenvalues, except for the one obtained in
section 3.

\subsubsection{Non-symmetric texture}

Until this point, we have implicitly taken the textures to be of
symmetric forms. Another more general analysis is to consider
non-symmetric types of texture. However, the general analysis is too
complex and is not particularly physically meaningful. Therefore we
suppose the up-type texture to be of the type A\@. We made numerical
analyses of the types of $4\times 4$ texture for the down-quark and
charged lepton sectors which reproduce the hierarchical mass ratios
charm/top, strange/bottom, mu/tau, etc., without the small parameter
$\epsilon$. Even in this situation, we found no realistic candidate,
except for the one obtained in section 3.

\subsection{The renormalization-group equations}
\setcounter{equation}{0}

We present the 2-loop beta-functions for gauge couplings of
$SU(3)_C\times SU(2)_W \times U(1)_Y$ and the 1-loop beta-functions
for Yukawa couplings. Here we neglect the $CP$ phase, which does not
affect the numerical results. The evolution of gauge coupling
constants is given by
\begin{eqnarray}
  \frac{d g_i}{d t} &=& b_i \frac{g_i^3}{16 \pi^2} + \frac{g_i^3}{(16
    \pi^2)^2} \left[ \sum_j b_{ij} g_j^2 - \sum_{a=u,d,e} c_{ia}
    \left( {\rm Tr} ({\mbox{\boldmath $Y$}}_a^T {\mbox{\boldmath
    $Y$}}_a) + Y_{\bar a}^2 \right) \right. \nonumber \\
  &&\hspace*{5cm}\left. - \sum_{k=1}^4 \sum_{X=Q,u,d,L,e}
    \hspace*{-2ex}d_{iX} Y_{X_k}^2 \right],
\end{eqnarray}
where  $b_i = (10.6,\,5,\,1)$ for $U(1)_Y$ (in a GUT normalization),
$SU(2)_W$ and $SU(3)_C$ respectively, and
\begin{eqnarray}
  b_{ij} &=& \pmatrix{
    977/75 & 39/5 & 88/3 \cr
     13/5  &  53  &  40  \cr
     11/3  &  15  & 178/3    }\,, \\
  c_{ia} &=& \bordermatrix{
    &   u  &   d  &   e  \cr
    & 26/5 & 14/5 & 18/5 \cr
    &   6  &   6  &  2   \cr
    &   4  &   4  &  0      }\,, \\
  d_{iX} &=& \bordermatrix{
    &  Q  &   u  &  d  &  L  &  e   \cr
    & 2/5 & 16/5 & 4/5 & 6/5 & 12/5 \cr
    &  6  &   0  &  0  &  2  &  0   \cr
    &  4  &   2  &  2  &  0  &  0     }\,.
\end{eqnarray}
The beta-functions for Yukawa couplings in the superpotential
(\ref{spot}) are given as follows:
\begin{eqnarray}
  \frac{d {\mbox{\boldmath $Y$}}_{\!a_{\,ij}}}{d t} &=& \frac{1}{16
  \pi^2} \beta_{a_{ij}}\,, \quad (a =u,d,e) \\
  \frac{d Y_{\bar a}}{d t} \;&=& \frac{1}{16\pi^2} \beta_{\bar a}\,,
  \quad\;\; (a =u,d,e) \\
  \frac{d Y_{X_i}}{d t} &=& \frac{1}{16 \pi^2} \beta_{X_i}\,,
  \quad\, (X=Q,u,d,L,e) \\
  \frac{d Y}{d t} \;&=& \frac{1}{16 \pi^2} \beta_Y,
\end{eqnarray}
\begin{eqnarray}
  &&\hspace*{-1cm} \beta_{u_{ij}} = {\mbox{\boldmath
      $Y$}}_{\!u_{\,ij}} \left[ 3 {\rm Tr} ({\mbox{\boldmath $Y$}}_u^T
      {\mbox{\boldmath $Y$}}_u) + 3 Y_{\bar d}^2 + Y_{\bar e}^2
      -\frac{16}{3} g_3^2 - 3 g_2^2 - \frac{13}{15} g_1^2 \right]
  \nonumber \\
  &&\hspace*{2mm} + \left( 3 {\mbox{\boldmath $Y$}}_u {\mbox{\boldmath
        $Y$}}_u^T {\mbox{\boldmath $Y$}}_u + {\bf Y}_u {\mbox{\boldmath
        $Y$}}_d^T {\mbox{\boldmath $Y$}}_d \right)_{ij} + \sum_k \left
    ( {\bf Y}_{\!u_{\,kj}} Y_{Q_i} Y_{Q_k} + {\mbox{\boldmath
        $Y$}}_{\!u_{\,ik}} Y_{u_k} Y_{u_j} \right), \\
  &&\hspace*{-1cm} \beta_{d_{ij}} = {\mbox{\boldmath
      $Y$}}_{\!d_{\,ij}} \left[ {\rm Tr} ( 3 {\mbox{\boldmath
        $Y$}}_d^T {\mbox{\boldmath $Y$}}_d + {\mbox{\boldmath $Y$}}_e^T
    {\mbox{\boldmath $Y$}}_e ) + 3 Y_{\bar u}^2 - \frac{16}{3} g_3^2
    - 3 g_2^2 - \frac{7}{15} g_1^2 \right] \nonumber \\
  &&\hspace*{2mm} + \left( 3 {\mbox{\boldmath $Y$}}_d {\mbox{\boldmath
        $Y$}}_d^T {\mbox{\boldmath $Y$}}_d + {\bf Y}_d {\mbox{\boldmath
        $Y$}}_u^T {\mbox{\boldmath $Y$}}_u \right)_{ij} + \sum_k \left
    ( {\bf Y}_{\!d_{\,kj}} Y_{Q_i} Y_{Q_k} + {\mbox{\boldmath
        $Y$}}_{\!d_{\,ik}} Y_{d_k} Y_{d_j} \right), \\
  &&\hspace*{-1cm} \beta_{e_{ij}} = {\mbox{\boldmath
      $Y$}}_{\!e_{\,ij}} \left[ {\rm Tr} ( 3 {\mbox{\boldmath
        $Y$}}_d^T {\mbox{\boldmath $Y$}}_d + {\mbox{\boldmath $Y$}}_e^T
    {\mbox{\boldmath $Y$}}_e ) + 3 Y_{\bar u}^2 - 3 g_2^2 -
    \frac{9}{5} g_1^2 \right] \nonumber \\
  &&\hspace*{2mm} + 3 \left( {\mbox{\boldmath $Y$}}_e {\mbox{\boldmath
        $Y$}}_e^T {\mbox{\boldmath $Y$}}_e \right)_{ij} + \sum_k \left
    ( {\mbox{\boldmath $Y$}}_{\!e_{\,kj}} Y_{L_i} Y_{L_k} +
    {\mbox{\boldmath $Y$}}_{\!e_{\,ik}} Y_{e_k} Y_{e_j} \right), 
\end{eqnarray}
\begin{eqnarray}
  \beta_{\bar u} &=& Y_{\bar u} \left[ {\rm Tr} ( 3 {\mbox{\boldmath
        $Y$}}_d^T {\bf Y}_d + {\mbox{\boldmath $Y$}}_e^T
    {\mbox{\boldmath $Y$}}_e ) + 6 Y_{\bar u}^2 + Y_{\bar d}^2 +\sum_i
    \left( Y_{Q_i}^2 + Y_{u_i}^2 \right) \right.\nonumber \\
  &&\biggl.\hspace*{3cm} - \frac{16}{3} g_3^2 - 3
    g_2^2 - \frac{13}{15} g_1^2 \biggr],\\ 
  \beta_{\bar d} &=& Y_{\bar d} \left[ {\rm Tr} ( 3 {\mbox{\boldmath
        $Y$}}_u^T {\bf Y}_u ) + Y_{\bar u}^2 + 6 Y_{\bar d}^2 +
    Y_{\bar e}^2 + \sum_i \left( Y_{Q_i}^2 + Y_{d_i}^2 \right)
  \right.\nonumber \\
  &&\biggl.\hspace*{3cm} -\frac{16}{3} g_3^2 - 3 g_2^2 - \frac{7}{15}
    g_1^2  \biggr],\\
  \beta_{\bar e} &=& Y_{\bar e} \left[ {\rm Tr} ( 3 {\mbox{\boldmath
        $Y$}}_u^T {\bf Y}_u ) + 3 Y_{\bar d}^2 + 4 Y_{\bar e}^2 +
    \sum_i \left ( Y_{L_i}^2 + Y_{e_i}^2 \right) - 3 g_2^2 -
    \frac{9}{5} g_1^2 \right],~~~ 
\end{eqnarray}
\begin{eqnarray}
  &&\hspace*{-1cm} \beta_{Q_i} = Y_{Q_i} \biggl[ Y_{\bar u}^2 +
  Y_{\bar d}^2 + \sum_j \left( 8 Y_{Q_j}^2 + 3 Y_{u_j}^2 + 3 Y_{d_j}^2
    + 2 Y_{L_j}^2 + Y_{e_j}^2 \right) + Y^2 \biggr. \nonumber \\
  &&\hspace*{1cm}\biggl.\qquad\; - \frac{16}{3} g_3^2 - 3 g_2^2 -
    \frac{1}{15} g_1^2 \biggr] + \sum_k Y_{Q_k} \left
    ( {\mbox{\boldmath $Y$}}_u^T {\mbox{\boldmath $Y$}}_u +
    {\mbox{\boldmath $Y$}}_d^T {\mbox{\boldmath $Y$}}_d
  \right)_{ik}\,,\\
  &&\hspace*{-1cm} \beta_{u_i} = Y_{u_i} \biggl[ 2 Y_{\bar u}^2 +
  \sum_j \left( 6 Y_{Q_j}^2 + 5 Y_{u_j}^2 + 3 Y_{d_j}^2 + 2 Y_{L_j}^2
    + Y_{e_j}^2 \right) + Y^2 \biggr. \nonumber \\
  &&\hspace*{1cm}\biggl.\qquad\; - \frac{16}{3} g_3^2 - \frac{16}{15}
  g_1^2 \biggr] + \sum_k 2 Y_{u_k} \left( {\mbox{\boldmath $Y$}}_u^T
    {\mbox{\boldmath $Y$}}_u \right)_{ik}\,, \\
  &&\hspace*{-1cm} \beta_{d_i} = Y_{d_i} \biggl[ 2 Y_{\bar d}^2 +
  \sum_j \left( 6 Y_{Q_j}^2 + 3 Y_{u_j}^2 + 5 Y_{d_j}^2 + 2 Y_{L_j}^2
    + Y_{e_j}^2 \right) + Y^2 \biggr. \nonumber \\ 
  &&\hspace*{1cm}\biggl.\qquad\; - \frac{16}{3} g_3^2 - \frac{4}{15}
  g_1^2 \biggr] + \sum_k 2 Y_{d_k} \left( {\mbox{\boldmath $Y$}}_d^T
    {\mbox{\boldmath $Y$}}_d \right)_{ik}\,, \\
  &&\hspace*{-1cm} \beta_{L_i} = Y_{L_i} \biggl[ Y_{\bar e}^2 + \sum_j
  \left( 6 Y_{Q_j}^2 + 3 Y_{u_j}^2 + 3 Y_{d_j}^2 + 4 Y_{L_j}^2 +
    Y_{e_j}^2 \right) + Y^2 \biggr. \nonumber \\ 
  &&\hspace*{1cm}\biggl.\qquad\; - 3 g_2^2 - \frac{3}{5} g_1^2 \biggr]
  + \sum_k Y_{L_k} \left( {\mbox{\boldmath $Y$}}_e^T {\mbox{\boldmath
        $Y$}}_e \right)_{ik}\,, \\ 
  &&\hspace*{-1cm} \beta_{e_i} = Y_{e_i} \biggl[ 2 Y_{\bar e}^2 +
  \sum_j \left( 6 Y_{Q_j}^2 + 3 Y_{u_j}^2 + 3 Y_{d_j}^2 + 2 Y_{L_j}^2
    + 3 Y_{e_j}^2 \right) + Y^2 \biggr. \nonumber \\
  &&\hspace*{1cm} \biggl.\qquad\; - \frac{12}{5} g_1^2 \biggr] +
  \sum_k 2 Y_{e_k} \left( {\mbox{\boldmath $Y$}}_e^T {\mbox{\boldmath
        $Y$}}_e \right)_{ik}\,, \\
  &&\hspace*{-1cm} \beta_Y = 3 Y \biggl[ \sum_i \left( 6 Y_{Q_i}^2 + 3
      Y_{u_i}^2 + 3 Y_{d_i}^2 + 2 Y_{L_i}^2 + Y_{e_i}^2 \right) + Y^2
  \biggr]\,.
\end{eqnarray}

\newpage
\setlength{\baselineskip}{15.5pt}

\end{document}